\documentclass[aip,jcp,reprint,graphicx,longbibliography]{revtex4-1}
\usepackage{graphicx}
\usepackage[hidelinks]{hyperref}
\usepackage{amsmath} 
\usepackage{amssymb}
\usepackage{mathrsfs}
\usepackage{color}
\draft 
\begin{document}

\title{Chirality transfer in lyotropic twist-bend nematics}

\author{Anna Ashkinazi}
\affiliation{Physical \& Theoretical Chemistry Laboratory, Department of Chemisty, University of Oxford, South Parks Road, Oxford OX1 3QZ, United Kingdom}
\author{Hemani Chhabra}%
\affiliation{Beckman Institute, University of Illinois Urbana-Champaign, IL, USA}
\author{Anouar El Moumane}
\affiliation{Physical \& Theoretical Chemistry Laboratory, Department of Chemisty, University of Oxford, South Parks Road, Oxford OX1 3QZ, United Kingdom}
\author{Maxime M.\ C.\ Tortora}
\affiliation{Department of Quantitative and Computational Biology, University of Southern California, Los Angeles, CA, USA}
\author{Jonathan P.\ K.\ Doye}
\email{jonathan.doye@chem.ox.ac.uk}
\affiliation{Physical \& Theoretical Chemistry Laboratory, Department of Chemisty, University of Oxford, South Parks Road, Oxford OX1 3QZ, United Kingdom}

\date{\today}

\begin{abstract}
Using molecular simulations and classical density functional theory, we study the liquid-crystalline phase behaviour of a series of bent rod-like mesogens with a controlled degree of chirality introduced through a twist at the centre of the particle. In the achiral limit, isotropic, uniaxial nematic, twist-bend nematic and smectic phases form as the packing fraction increases. On introducing chirality, the symmetry between the right- and left-handed twist-bend phases is broken. The phase with the same-handedness as the particles quickly becomes overwhelmingly favoured as the magnitude of the particle twist is increased, because the particles are then able to better follow the helical director field lines in the twist-bend phase and pack more efficiently. By contrast, the cholesteric phase is predicted to have the opposite handedness to that of the particle due to the relatively weakly-twisted nature of the particles. That the cholesteric and twist-bend phases have opposite handedness illustrates the differences in the mechanisms of chirality transfer in the two phases. 
We also found that doping a system of achiral mesogens with a small fraction of chiral particles led to selection of the twist-bend phase with the same chirality as the particle.
\end{abstract}


\maketitle 

\section{Introduction}
\label{sect:intro}

Shape anisotropy plays a key role in the stabilization of liquid-crystalline phases. The simplest examples are uniaxial rods, which typically exhibit nematic and smectic liquid-crystalline phases where the orientational order in these phases allows the particles to pack more efficiently. If the particle shape exhibits additional features, the nematic phase can exhibit deviations from uniaxial order. For example, chiral rods exhibit cholesteric (or chiral nematic) phases. More recently, it has been found that bent mesogens can exhibit modulated nematic phases, such as the twist-bend nematic.\cite{Cestari11,Chen13}

\begin{figure}
\includegraphics[width=4cm]{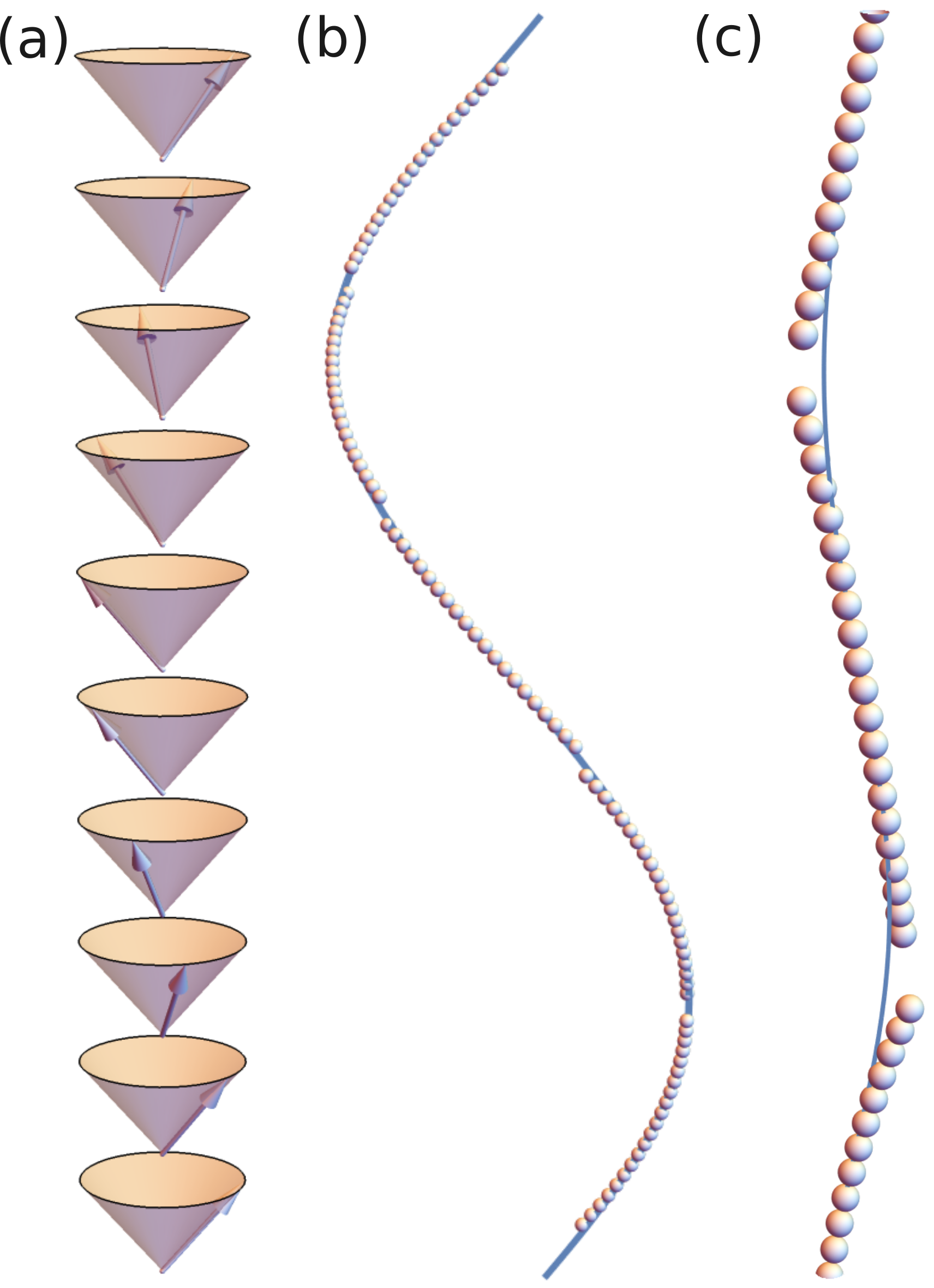}
\caption{\label{fig:TB_schematic}(a) The variation of the nematic director (arrows) associated with the heliconical order in the twist-bend phase. The angle between the local nematic director and the direction of modulation is constant and is termed the cone angle $\theta$. (b) and (c) A helical path of a nematic directory field line decorated with achiral bent particles showing how the particles are able to approximately follow the director field. In all cases the sense of the rotation of the nematic director is left-handed and one pitch length has been visualized. The pitch $\mathcal{P}$ and cone angle $\theta$ are in (a) and (b) $89.5\,\sigma$ and $40.43^\circ$ and in (c) $39.6\,\sigma$ and $10.98^\circ$. These have been chosen to match the values for the $\alpha$=0 and $\alpha$=$-90^\circ$ particles, respectively,
at $p$=0.05\,$\epsilon\sigma^{-3}$. In (c) the deviations of the achiral particles from the field lines are more significant due to the latter's greater torsion.}
\end{figure}

\begin{figure}
\includegraphics[width=6cm]{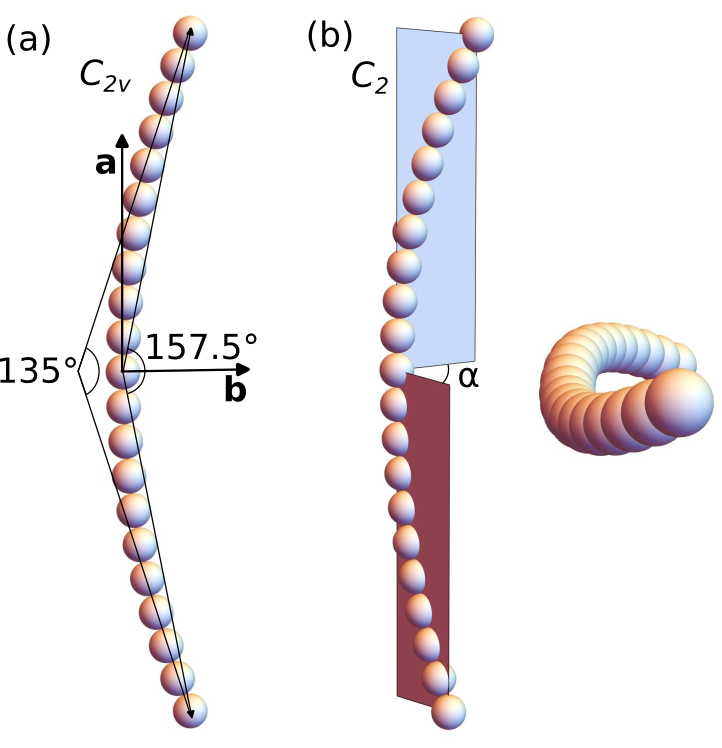}
\caption{\label{fig:particle}(a) Model achiral mesogen consisting of 21 tangent spheres placed on an arc with a central angle of 45$^\circ$. The molecular axis $\mathbf{\hat{a}}$ is defined as the tangent to the arc at the centre of the particle. All the spheres lie on a plane that corresponds to a mirror plane of the system. 
The secondary $\mathbf{\hat{b}}$ axis is perpendicular to $\mathbf{\hat{a}}$ and in the plane of the particle.
(b) A model chiral mesogen derived from the achiral mesogen by rotating the two halves of the particle by an angle $\alpha/2$ about the molecular axis but in opposite directions. The angle between the planes on which the two halves lie is $\alpha$. The illustrated example has $\alpha$=90$^\circ$, is right-handed (as is clear from the second view) and the curvature of the two halves of the particle has been adjusted so that the angle between the two midpoint-to-end vectors is the same as that for the achiral particle, namely 157.5$^\circ$.
}
\end{figure}

In the twist-bend phase, the director field 
$\mathbf{\hat{n}}(\mathbf{r})$ 
exhibits heliconical order (Fig.\ \ref{fig:TB_schematic}(a)); i.e.\ 
$\mathbf{\hat{n}}(\mathbf{r})=
\left(\sin\theta \cos(2\pi z/\mathcal{P}), 
\sin\theta \sin(2\pi z/\mathcal{P}),
\cos\theta \right)$. 
The two key geometric characteristics of the phase are the cone angle $\theta$, the angle between the local nematic director and its average, and the pitch $\mathcal{P}$, the length scale over which the nematic director field completes a full rotation.
The phase results from the development of additional polar order associated with the secondary $\mathbf{\hat{b}}$ axis of the molecule (Fig.\ \ref{fig:particle}), driven by the better packing that results if the bent particles ``cup'' into each other.
The twist-bend phase was first predicted based on theoretical arguments on the effect of this polar order on the bend elastic constant of the nematic phase.\cite{Dozov01} In particular, if this elastic constant becomes negative, the uniaxial nematic becomes unstable with respect to bending; the twist-bend phase represents one way to incorporate bending of the director field into a nematic phase (another possibility is the splay-bend nematic).

The director field lines in the twist-bend phase are helical. Fig.\ 1(b) and (c) illustrate for a single field line how achiral bent particles can approximately follow the director field and, thus, provide a model for the organization of the particles in the twist-bend phase. Of course, packing the particles in this manner in a bulk system is going to be more complicated than around a single axis, particularly as the packing of twisted objects is often frustrated.\cite{Grason16}

Twist-bend nematics were first discovered in thermotropic liquid crystals and associated with mesogens with a ``bent-core'' geometry.\cite{Cestari11,Chen13} The molecular characteristics of the nematogens that are required to exhibit a twist-bend phase have been extensively explored.\cite{Mandle22}
The tendency to form twist-bend phases has also been reproduced in simulations of archetypal molecules.\cite{Yu22,Wilson23}

Simulations of simple model particles can also be used to explore the minimal requirements for a twist-bend phase to be observed. These have mainly focussed on lyotropic systems of bent rods,\cite{Greco15,Chiappini19,Chiappini21,Kubala22,Subert24,Vanakaras25,Birdi25} where the interactions are purely repulsive and the phase transitions are driven by increasing particle concentration. For smoothly bent rods, twist-bend phases are observed at intermediate degrees of bending, but for rods with a sharp bend in the middle of the particle, smectic ordering can often be more favourable.\cite{Chiappini19}

Recently, methods to synthesise bent colloidal rods have been developed, and the phase behaviour of the resulting particles studied.\cite{FernandezRico20,Kotni22,FernandezRico21,FernandezRico24} Modulated nematic phases have been observed, albeit with a preference for splay-bend rather than twist-bend order that is perhaps due to the combined effects of gravity and the flat wall at the bottom of the sample.\cite{FernandezRico20,Kotni22}

A variety of theoretical approaches have been developed to describe the twist-bend nematic phase. These include Landau-de Gennes approaches applied to thermotropics \cite{Longa20} and lyotropics \cite{Anzivino20} and generalized Maier-Saupe frameworks.\cite{Chiappini21,Revignas24} 
Second-virial density functional theories have also been used to study the effects of increasing concentration on the nematic elastic constants of hard bent particles, observing the softening of the bend elastic constant that leads to spontaneous bend deformations.\cite{DeGregorio16,Revignas20,Revignas21,Revignas22}

In twist-bend phases, the heliconical order can be right- or left-handed. For achiral particles, the expectation is that each handedness will occur 50\% of the time. Indeed, for most thermotropic systems that form a twist-bend phase, no handedness preference is observed. However, chiral twist-bend phases have been identified in a limited number of cases.\cite{Gorecka15,Walker19,Walker20,Walker23,Ozegovic24,Ozegovic24b} Chiral phases can also be achieved by the addition of a chiral dopant to a twist-bend forming system.\cite{Archbold15,Murachver19,Zhang22b}

A molecular understanding of how particle chirality determines phase chirality in lyotropic cholesteric phases has been broadly outlined. For screw-like objects Straley's geometric principle applies;\cite{Straley76} namely, the phase handedness is expected to reflect the handedness of the configuration for which the threads of two particles interlock on close approach. Thus, for weakly twisted particles, the cholesteric phase is expected to have the opposite handedness to the particles, whereas for strongly twisted particles the particle and phase are expected to have the same handedness;
a thread angle of $45^\circ$ separates these two regimes. 
This principle has been shown to apply well for simple rigid chiral particles, albeit with the crossover between handedness occurring near to, rather than at, 45$^\circ$.\cite{Frezza14,Tortora17b} For more complex chiral particles where the application of the above geometric argument may not be straightforward, density functional theory can be used to compute the chiral strength $k_t$ (its sign determines the handedness of the cholesteric phase),\cite{Dussi15,Tortora17} including the effects of thermal fluctuations for non-rigid particles.\cite{Tortora18} This approach has been applied to resolve the origin of the handedness in chiral DNA origami rods \cite{Tortora20} and filamentous viruses.\cite{Grelet24}

By contrast, there has been relatively little consideration of chirality transfer for the twist-bend phase. The theories that have considered the chiral twist-bend phase have done so just by the introduction of a chiral strength term that is linear in the twist.\cite{Meyer16,Longa18}
Hence, the sign of this term determines the predicted handedness of both the cholesteric and the twist-bend phases. Although it is well established that this term is the main factor for the cholesteric, this is less plausible for the twist-bend phase. Rather, it might be expected that the coupling of the particle chirality to the director deformations in the twist-bend phase would play an important role. For example, such couplings between particle morphology and director distortions have been shown to significantly affect the nematic elastic constants;\cite{DeGregorio16,Revignas20,Revignas21,Revignas22} it is this coupling that leads to the softening of the bend elastic constant and the onset of twist-bend order for bent particles.

Here, we use simulations to study chiral bent lyotropic mesogens to begin to address the question of chirality transfer in the twist-bend phase. A particular focus is on how the handedness of the phase depends on the particle handedness and how the properties of the right- and left-handed twist-bend phase bifurcate as increasing particle chirality is introduced.

\section{Methods}
\label{sect:methods}

\subsection{Model particles}
\label{sect:particles}

The achiral bent particle that is the starting point for our study is illustrated in Fig.\ \ref{fig:particle}(a). It consists of 21 tangent spheres whose centres are placed on an arc with central angle 45$^\circ$. Thus, the angle between the tangents to the arc at the end spheres is 135$^\circ$. 
We chose this value based on the phase diagram computed for similar achiral mesogens (albeit at lower aspect ratio)\cite{Kubala22} and
because preliminary simulations indicated robust formation of the twist-bend nematic phase. 
A bend angle can also be defined as the angle between the vectors connecting the central sphere to the end spheres; this angle is 157.5$^\circ$. 
Very similar mesogens have been studied by simulation 
\cite{Greco15,Chiappini19,Chiappini21,Kubala22,Subert24,Vanakaras25,Birdi25} 
and theory.\cite{Greco15,Chiappini21,DeGregorio16,Revignas20,Revignas22,Revignas24} 

Our aim is to study a series of particles that have a controlled degree of chirality but similar bending and that have the achiral bent particle as one limit.
To generate such a particle 
we choose to rotate the top and bottom halves of the achiral particle by an angle of $\alpha/2$ 
about the principal axis, but in opposite senses, as illustrated in Fig.\ \ref{fig:particle}(b). The illustrated example is right-handed as is clear from the end view of the particle. A left-handed particle can be obtained by rotations in the opposite sense. 
The convention that we use is to assign a negative value of $\alpha$ to left-handed particles.

The achiral particle is based on a curve that has constant curvature and zero torsion. 
A helix has constant curvature and torsion. 
The chiral particles have constant curvature and zero torsion at all points except the centre of the particle, where the torsion exhibits a delta function.
An alternative option might have been to consider a series of particles based on a helix with the curvature kept constant and the torsion $\tau$ as a parameter ($\tau$=0 would correspond to the achiral limit). 
The total twist in such particles would be $\alpha=\tau L$ where $L$ is the length of the particle; for a given $\alpha$ our particles are effectively more chiral as introducing all the twist at the centre has a greater effect on the particle shape.

Our definition of the molecular axis of the particle (as the tangent to the midpoint of the particle) remains unchanged by the addition of chirality. We note, though, that it is no longer parallel to the end-to-end vector of the particle or the eigenvector of the gyration tensor with the largest eigenvalue; either of these could also have been used but the choice has only a small effect on quantities such as the nematic order parameter. 
Furthermore, as $\alpha$ increases, the bend angle defined by the vectors from the central to the end spheres of the particle increases if the curvature is unchanged. The particles effectively become somewhat less bent. For example, for $\alpha$=90$^\circ$ this angle would increase to 164.1$^\circ$ from 157.5$^\circ$ for $\alpha$=0. To compensate for the effect of $\alpha$ on the bent character of the particles, we increase the curvature as $\alpha$ is increased such that this bend angle is kept fixed at 157.5$^\circ$.

We consider values of $\alpha$ in the range 0--90$^\circ$. Beyond 90$^\circ$ the chirality reduces, eventually giving an achiral $C_{2v}$ S-shaped particle at $\alpha$=180$^\circ$.

The mesogens are rigid. Spheres in different particles interact via a purely repulsive WCA (Weeks-Chandler-Andersen) potential;\cite{Weeks71} i.e.\ the interaction is a Lennard-Jones potential:
\begin{equation}
V_\mathrm{LJ}(r) = 4\varepsilon{}\left[\left( \frac{ \sigma}{r} \right)^{12} - \left( \frac{\sigma{}}{r} \right)^6 \right]
\label{eq:LJ}
\end{equation}
that has been cut and shifted so that the potential goes smoothly to zero at $2^{1/6}\sigma$; i.e.\ 
\begin{equation}
V_\mathrm{WCA}(r)=
\begin{cases}
V_\mathrm{LJ}(r) + \varepsilon & : r<2^{1/6}\sigma \\
0 & : r\ge 2^{1/6}\sigma,
\end{cases}
\end{equation}
where \(r\) is the separation between two spheres. We use \(\varepsilon\) and \(\sigma\) as our units of energy and length.
The separation of adjacent spheres within a particle is $\sigma$.

\subsection{Order parameters}
\label{sect:analysis}

The average nematic director was simply calculated by
\begin{equation}
\mathbf{\hat{n}}=\frac{1}{N}\sum_{i=1}^{N} \mathbf{\hat{a}}_i.
\end{equation}
Although at equilibrium the above sum should be zero because configurations with $\mathbf{\hat{a}}_i$ parallel or anti-parallel to $\mathbf{\hat{n}}$ are equivalent, in our case this definition works fine because head-to-tail flipping of the mesogens in the nematic phases is slow compared to our simulation time scales and memory of the initial configuration in which are all $\mathbf{\hat{a}}_i$ are parallel is retained. 
A nematic order parameter 
$S$ 
can then be calculated as
\begin{equation}
S=\frac{1}{N}\sum_{i=1}^{N} \frac{3}{2}\left(\mathbf{\hat{a}_i} \cdot \mathbf{\hat{n}} -\frac{1}{3}\right).
\end{equation}
Note, we have defined this order parameter with respect to the average nematic director rather than the local director field.

In the twist-bend nematic phase, the nematic director has a heliconical spatial modulation. To characterize the twist-bend phase, we computed the local nematic director in slices of thickness $2\,\sigma$ perpendicular to the average nematic director. When a well-ordered twist-bend nematic phase was formed, it was a simple matter to obtain the pitch $\mathcal{P}$, the cone angle $\theta$ and the handedness of the twist-bend phase from the variations of this local nematic director.

To aid the identification of the smectic phase we also calculated the smectic order parameter $\Sigma$: 
\begin{equation} 
\Sigma = \frac{1}{N} \sum_{j=1}^{N} \exp{ \left( i\mathbf{k} \cdot \mathbf{r}_j \right)},
\label{eq:smectic}
\end{equation}
where $\mathbf{r}_j$ is the position of particle $j$ and $\mathbf{k}$ is the wavevector associated with the density modulation. 

\begin{figure*}
\includegraphics[width=17.4cm]{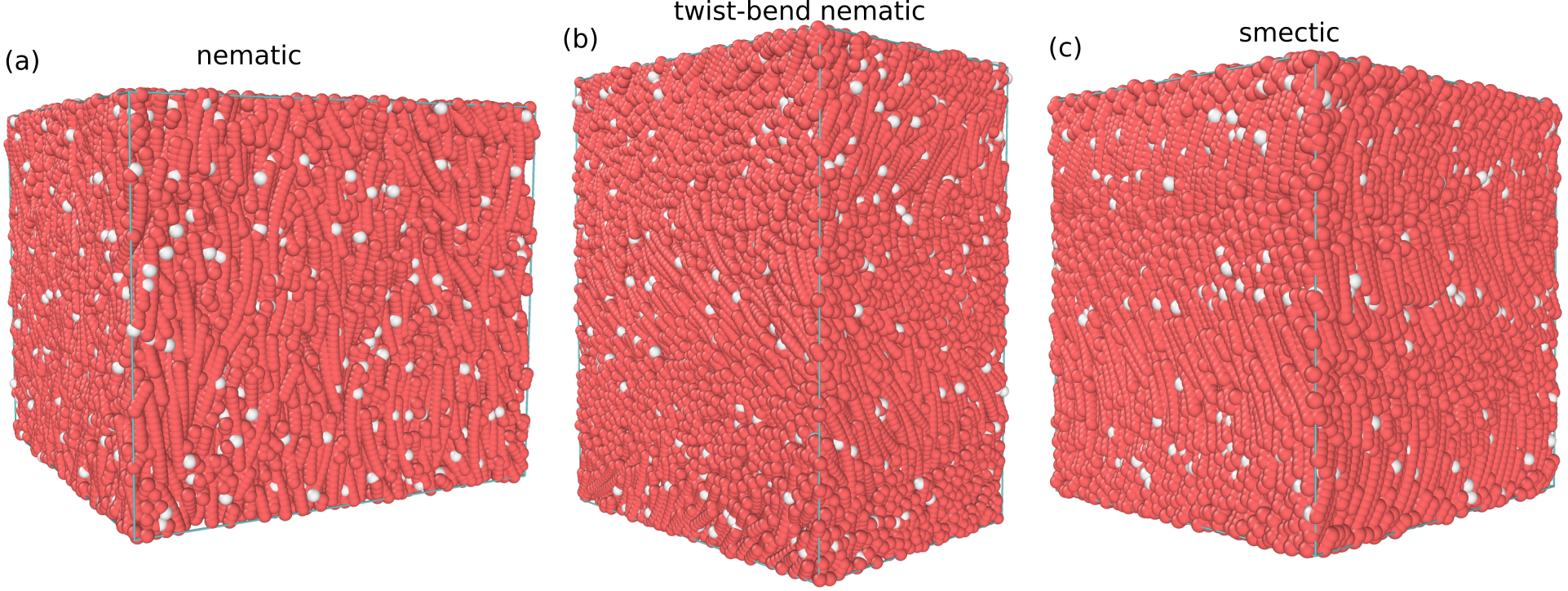}
\caption{\label{fig:phases} Example liquid-crystalline phases for the achiral ($\alpha$=0) bent mesogen: 
(a) uniaxial nematic ($N$) ($\eta$=0.242), 
(b) twist-bend nematic ($N_\mathrm{TB}$) ($\eta$=0.294); the box length in $z$ corresponds to one pitch length, (c) Smectic ($Sm$) ($\eta$=0.411).
The middle sphere in a particle is white to aid the identification of smectic order or its absence. 
}
\end{figure*}

\subsection{Simulations}
\label{sect:sims}

Molecular dynamics simulations were performed using the HOOMD-blue code \cite{Anderson20} in the $NPT$ ensemble. In these repulsive lyotropic systems, the behaviour is expected to mainly depend on the ratio $P/k_BT$. We used a fixed temperature $T$=0.2\,$\epsilon k_B^{-1}$ 
and varied $P$ to access the different phases. However, we report most of our results as a function of the packing fraction.
As our particle is designed to approximate a smoothly bent spherocylinder, we define the packing fraction using the volume of this spherocylinder as the particle volume, i.e.\
$20 \sigma \pi \left(\sigma/2\right)^2 + (4/3)\pi \left(\sigma/2\right)^3$, 
as has been done previously.\cite{Subert24}
All simulations used 6000 particles. 

The systems are initialized with the particles on a low-density primitive orthorhombic lattice all identically oriented (molecular axis parallel to $z$), thus ensuring no particle overlaps. The initial rapid compression at the start of the simulation leads to the rapid loss of lattice order. Although some smectic order typically remains, this is usually also soon lost (except for simulations in those parts of state space where a smectic is the stable phase).

Identification of the phase is usually straightforward, both from inspecting configurations (Fig.\ \ref{fig:phases}) and from the order parameters defined in the previous section.  The equations of state also show discontinuities at the phase boundaries (Fig.\ S1).
Occasionally, although the tendency towards forming a modulated nematic phase is apparent, the heliconical order is not sufficiently established to definitively assign the handedness (mainly at low $\alpha$). These are not included in the statistics for handedness.

\begin{figure}
\includegraphics[width=8.4cm]{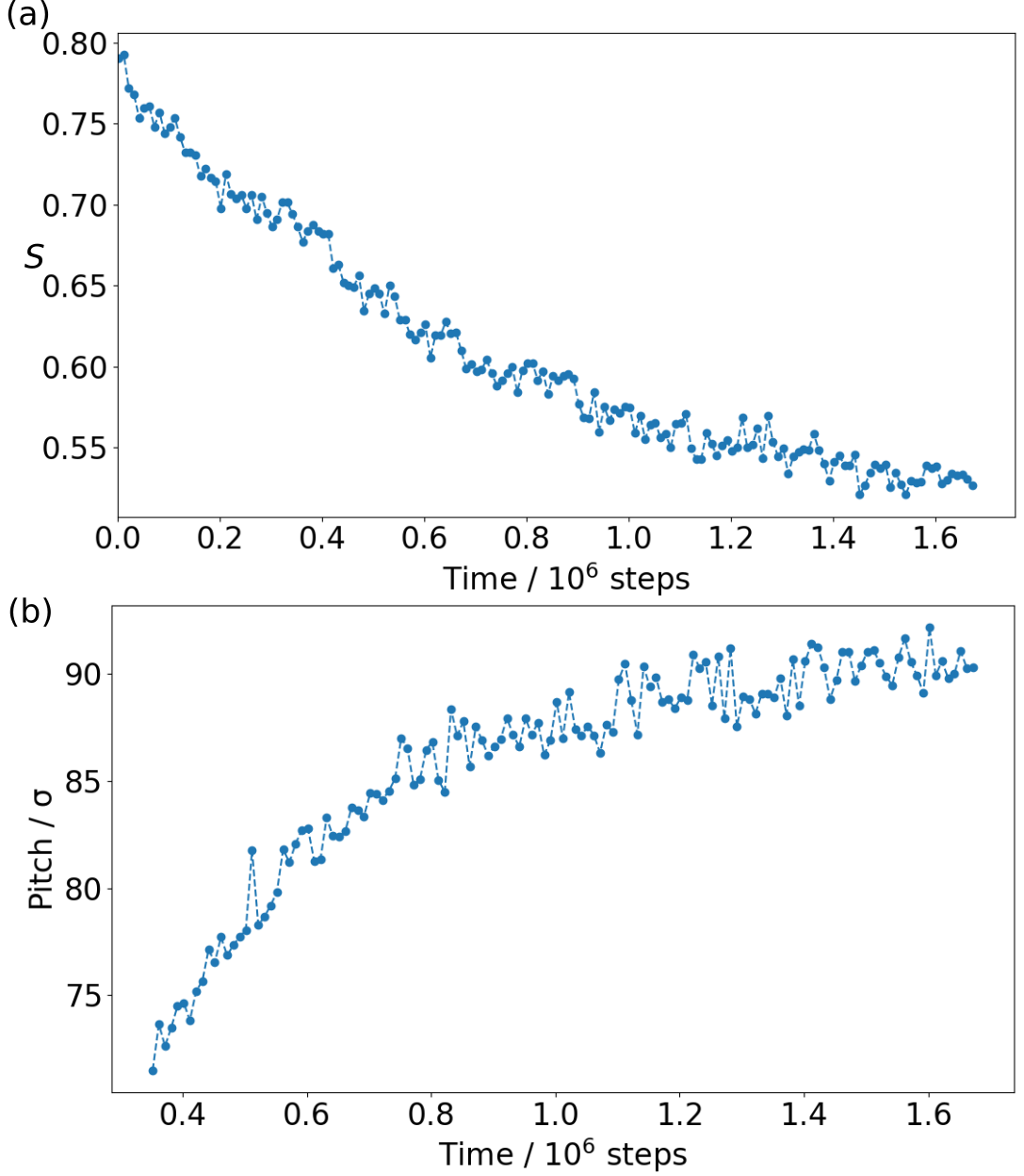}
\caption{\label{fig:equil} An example of the equilibration of the properties of the nematic twist-bend phase: (a) nematic order parameter $S$
and (b) pitch length of the twist-bend nematic phase versus time.
$\alpha$=0 and pressure $P=0.05\,\epsilon\sigma^{-3}$.
}
\end{figure}

When the twist-bend nematic phase forms, due to the initial configuration, the average nematic director is usually (but not always) oriented along $z$ and the pitch must then be commensurate with the box length along this direction. Often, as in Fig.\ \ref{fig:phases} the box length is equal to a single pitch length. However, this is unlikely to correspond to the equilibrium pitch length. Relaxation to equilibrium is consequently relatively slow as it must involve a change in the aspect ratio of the box and accompanying mass transport. Fig.\ \ref{fig:equil} illustrates the equilibration of the pitch length and the nematic order parameter. At the end of our standard simulation time, the properties of the twist-bend phase are close to their equilibrium values.

We observe smectic phases at the highest packing fractions that we consider. However, they are not the focus of the paper, and so we have not attempted to investigate the precise nature of the ordering nor tried to ensure their properties are fully equilibrated---as the densest phase, their relaxation times are likely to be the longest. Previously, splay-bend smectics and twist-bend-splay smectics have been identified for similar bent particles,\cite{Chiappini21,Kubala22} and it is clear from Fig.\ \ref{fig:phases}(c) that modulations in the orientational order are present.

For $\alpha\neq 0$, the nematic phase is expected to be chiral (i.e.\ a cholesteric phase). However, as the cholesteric pitch is usually large, simulating cholesterics phases is challenging and typically requires large systems to be simulated (e.g.\ if periodic boundary conditions are used one of the box lengths would need to be at least half a pitch length).\cite{Dussi16} 
An alternative way to estimate the properties of the cholesteric phase is to use classical density functional,\cite{Tortora17} as outlined in the next section.

\subsection{Classical density functional theory}

The bulk free energy of a weakly distorted uniaxial nematic phase can be described at the continuum level by the standard Oseen-Frank field functional\cite{Frank58}
\begin{equation}
\begin{split}
\label{eq:frank}
\mathscr{F} = &\:\mathscr{F}_0 + \frac{1}{2} \int_V d\mathbf{r} \times \Big\{ K_1 \left(\nabla\cdot\mathbf{n}\right)^2 + K_2 \left(\mathbf{n} \cdot \left[\nabla\times\mathbf{n}\right]\right)^2 \\
                     &+ K_3 \left(\mathbf{n}\times\left[\nabla\times\mathbf{n}\right]\right)^2 + 2k_t\left(\mathbf{n} \cdot \left[\nabla\times\mathbf{n}\right]\right)\Big\}
\end{split}
\end{equation}
with 
$V$ the total volume of the sample. $\mathscr{F}_0$ corresponds to the free energy density of the uniformly aligned nematic phase, and $k_t$ is a chiral strength parameter which vanishes in the case of achiral mesogens. $K_1$, $K_2$ and $K_3$ are referred to as the {Frank elastic constants}, and quantify the respective free-energy penalties associated with splay, twist and bend deformation modes of the local director field.\cite{deGennes93} 
It should be noted that this compact expression derives from a finite-order truncation of the free-energy gradient expansion, and is therefore only valid in the limit of long-wavelength director fluctuations. 

For $k_t=0$ and positive $K_1$, $K_2$ and $K_3$, the uniaxial nematic phase is the free-energy global minimum of the above expression. However, for non-zero $k_t$, the global minimum is instead a cholesteric phase with pitch $2\pi K_2/k_t$. To estimate the properties of the cholesteric phase of our chiral bent particles, we follow the perturbative density functional theory approach used in Ref. \onlinecite{Tortora17}. This approach assumes weak phase chirality; i.e.\ the pitch is sufficiently large that the orientational distribution function in the uniaxial nematic and cholesteric phases can be assumed to be the same. Another approximation is the use of the Parsons-Lee approach\cite{Parsons79,Lee87b} to estimate the direct correlation function. The relevant virial-like integrals are computed using Monte Carlo sampling and the eigenvector of the gyration tensor with largest eigenvalue is used as the definition of the long axis $\mathbf{\hat{a}}$.

Due to the above approximations, the predicted values of the properties of the cholesteric phase, e.g.\ the pitch, should be interpreted with suitable caution. However, our main focus is on the handedness (i.e.\ the sign of $k_t$) and this approach should be sufficient for this to be correctly determined.
We note that for bent particles in particular, the uniaxial assumption is expected to become worse as the packing fraction increases due to the onset of polar and biaxial order in the orientational distribution function. 
Consequently, the bend and twist elastic constants are expected to be increasingly overestimated. More sophisticated non-local density functional theory approaches that take into account this additional order observe a decrease in bend and twist elastic constants (the latter particularly in presence of chirality) at higher packing fractions.\cite{DeGregorio16,Revignas20,Revignas22} 
The effects of these changes on the cholesteric phase as the twist-bend phase boundary is approached are interesting but beyond the scope of the current paper.

\section{Results}
\label{sect:results}

\subsection{1-component simulations}
\label{sect:onecomponent}

\begin{figure}
\includegraphics[width=8.4cm]{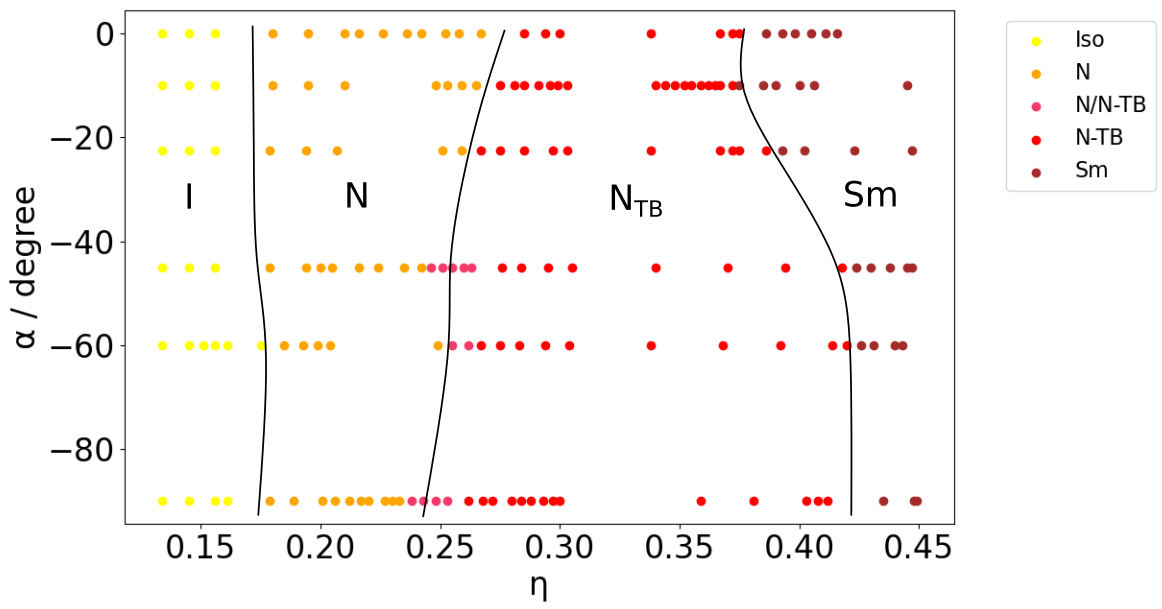}
\caption{\label{fig:phased} Phase diagram as a function of chiral angle $\alpha$ and packing fraction $\eta$. 
The points are coloured by the phase observed in the majority of the simulations at that state point. For some state points on the $N$/$N_\mathrm{TB}$ boundary, although there seemed to be a tendency to director modulations, clear heliconical was not yet present. The lines are guides to the eye.
}
\end{figure}

Figure \ref{fig:phased} shows the observed phases as a function of packing fraction $\eta$ and chiral angle $\alpha$. Note that the boundaries drawn are approximate given that they are based on the observed phase rather than rigorous calculations of the relative free energies of phases. The phase diagram is given for a single temperature, but in these lyotropic systems there is not expected to be much variation in the phase behaviour with temperature.

For the achiral particles the sequence of phases recapitulates what has been previously observed for smoothly bent particles of sufficient aspect ratio.  \cite{Greco15,Chiappini19,Chiappini21,Kubala22,Subert24} Namely, the system transitions from an isotropic (iso) phase to a uniaxial nematic ($N$) phase to a twist-bend nematic phase ($N_\mathrm{TB}$) and then finally to a smectic phase as the packing fraction increases. Examples of the liquid-crystalline phases are depicted in Fig.\ \ref{fig:phases}.

For these achiral particles, the nematic order parameter exhibits a step down at the $N$-$N_\mathrm{TB}$ transition, due to the tilting of the local nematic director in the twist-bend phase. 
As the packing fraction increases, the cone angle of the twist-bend phase increases somewhat (leading to a continued decrease in the nematic order parameter) and the pitch decreases. 
This is consistent with previous simulations \cite{Kubala22} and a Landau-de Gennes theory designed for lyotropic bent particles.\cite{Anzivino20} 

From the pitch and cone angle, the magnitude of the curvature $\kappa$ and torsion $\tau$ of the helical nematic director field lines can be computed; namely,
\begin{equation}
\kappa=2\pi\sin\theta\cos\theta/\mathcal{P}
\label{eq:curvature}
\end{equation}
and 
\begin{equation}
\tau=2\pi\cos^2\theta/\mathcal{P}. 
\label{eq:torsion}
\end{equation}
For example, at $P$=$0.05\,\epsilon\sigma^{-3}$ $1/\kappa$=$28.85\,\sigma$. This is close to the radius of curvature of the achiral particles, namely, 25.46\,$\sigma$. 
The curvature of the nematic director field approximately matches that of the particle, in agreement with the schematic picture of the order in the twist-bend phase in Fig.\ \ref{fig:TB_schematic}(b).

The main effect of increasing $|\alpha|$ on the phase diagram is to stabilize the twist-bend phase, significantly widening the range of packing fraction for which the twist-bend phase is observed. 
We should note that in the presence of chirality the uniaxial nematic phase is expected to transform into a cholesteric phase; however, in our simulations a uniaxial phase is still observed since the system sizes are not sufficiently large to allow the system to lower its free energy by forming a cholesteric phase. This constraint may have some small effect on the position of the $N$-$N_\mathrm{TB}$ boundary, but we should also note that when the transition from a uniaxial nematic to a twist-bend nematic occurs in a simulation, it is typically not directly to the lowest free-energy form of the twist-bend nematic due to the constraints of what the box size happens to be at the time of the transition (e.g.\ Fig.\ \ref{fig:equil}(b)).

As expected, there is relatively little effect of $\alpha$ on the phase boundary between the isotropic and nematic phases, as this is mainly controlled by the aspect ratio of the particles.
Although we observe the transition to the smectic phase to be pushed to higher packing fractions, perhaps because the chirality is less naturally incorporated into the smectic than the twist-bend phase, this is the phase boundary about which one should be most cautious due to the greater difficulty of equilibration at higher packing fractions.

For the left-handed mesogens that we consider, the twist-bend phase exhibits a clear tendency to form a left-handed phase. Thus, there is a strong preference for the particles and the twist-bend phase to have the same chirality. 
Why is this? The likely reason is that if the phase has the same handedness as the particle, it better allows the particles to follow the local nematic director field. For example, in Fig.\ \ref{fig:TB_schematic}(b) and (c) the achiral particles deviate somewhat from the helical paths of the field lines due to the torsion of the latter. 
Introducing a twist into the particles that has the same handedness as the phase allows these deviations to be reduced. By contrast, if the particle has a twist with the opposite handedness to the phase, these deviations will increase.
This better match between the particle and director modulations is also the likely reason for the stabilization of the twist-bend phase with increasing chirality.

To look more closely at the onset of a preferred chirality to the twist-bend phase, we studied the relative propensity to form the left- or right-handed phases as a function of $\alpha$ at a pressure near to the centre of the stability region of the twist-bend phase. Specifically, we ran 40 simulations 
at each value of $\alpha$ considered, measuring the fraction that had resulted in the right- or left-handed phase at the end of the simulation. 
A small fraction of the simulations, mainly close to $\alpha$=0, had not yet developed a sufficiently clear handedness at the end of the simulation and so were not included in the statistics.

Reassuringly, at $\alpha$=0 each handedness was equally likely (within statistical error), as expected. However, the onset of chiral phase selection is rapid as $|\alpha|$ increases and by $\alpha$=$-20^\circ$ the left-handed phase was favoured in the vast majority of the simulations. As a sanity check, we also tested that the results are mirrored for right-handed particles by simulating a particle with $\alpha=+45^\circ$; a right-handed phase always resulted.

\begin{figure}
\includegraphics[width=8.4cm]{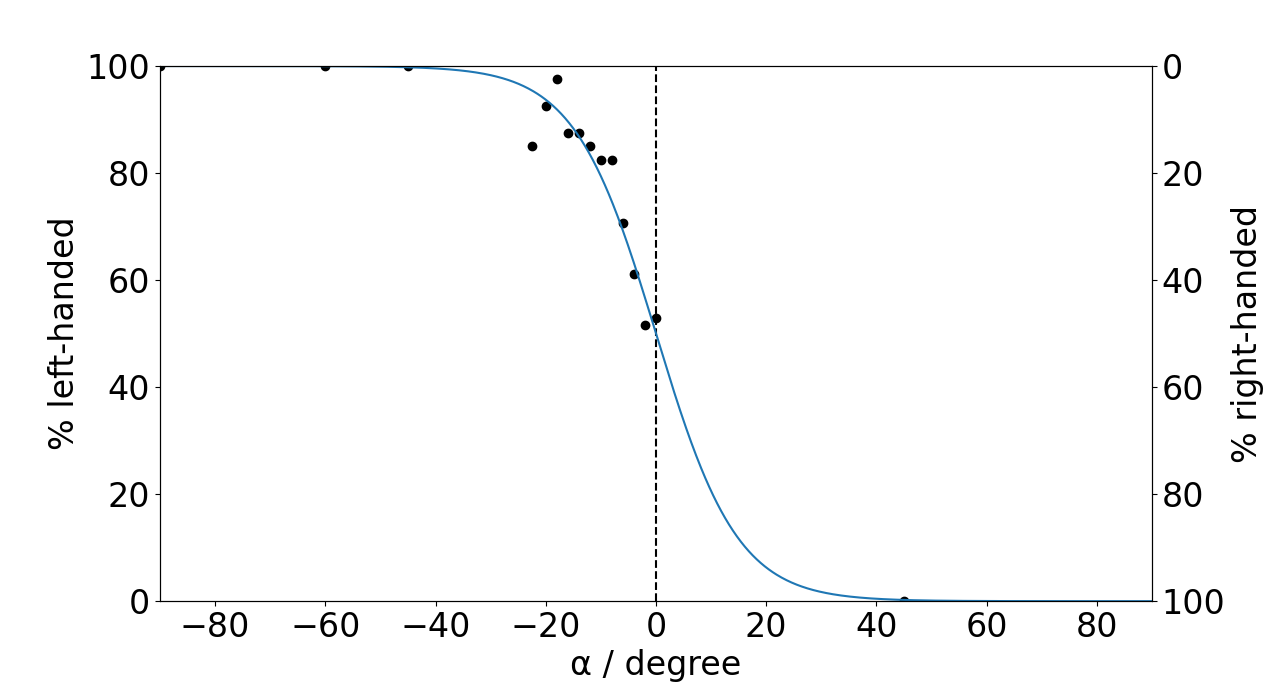}
\caption{\label{fig:LR} Probability of the twist-bend nematic phase being left- or right-handed  as a function of $\alpha$ at pressure $P$=$0.05\,\epsilon\sigma^{-3}$. The line is a fit (a $\tanh$ function) assuming anti-symmetry about $\alpha=0$.
}
\end{figure}

We also investigated how the properties of the right-handed and left-handed phases differed. As expected, at $\alpha$=0, properties such as pitch, cone angle, packing fraction and potential energy are essentially independent of the handedness of the phase. However, for non-zero $\alpha$ their properties diverge. 

\begin{figure}
\includegraphics[width=8.4cm]{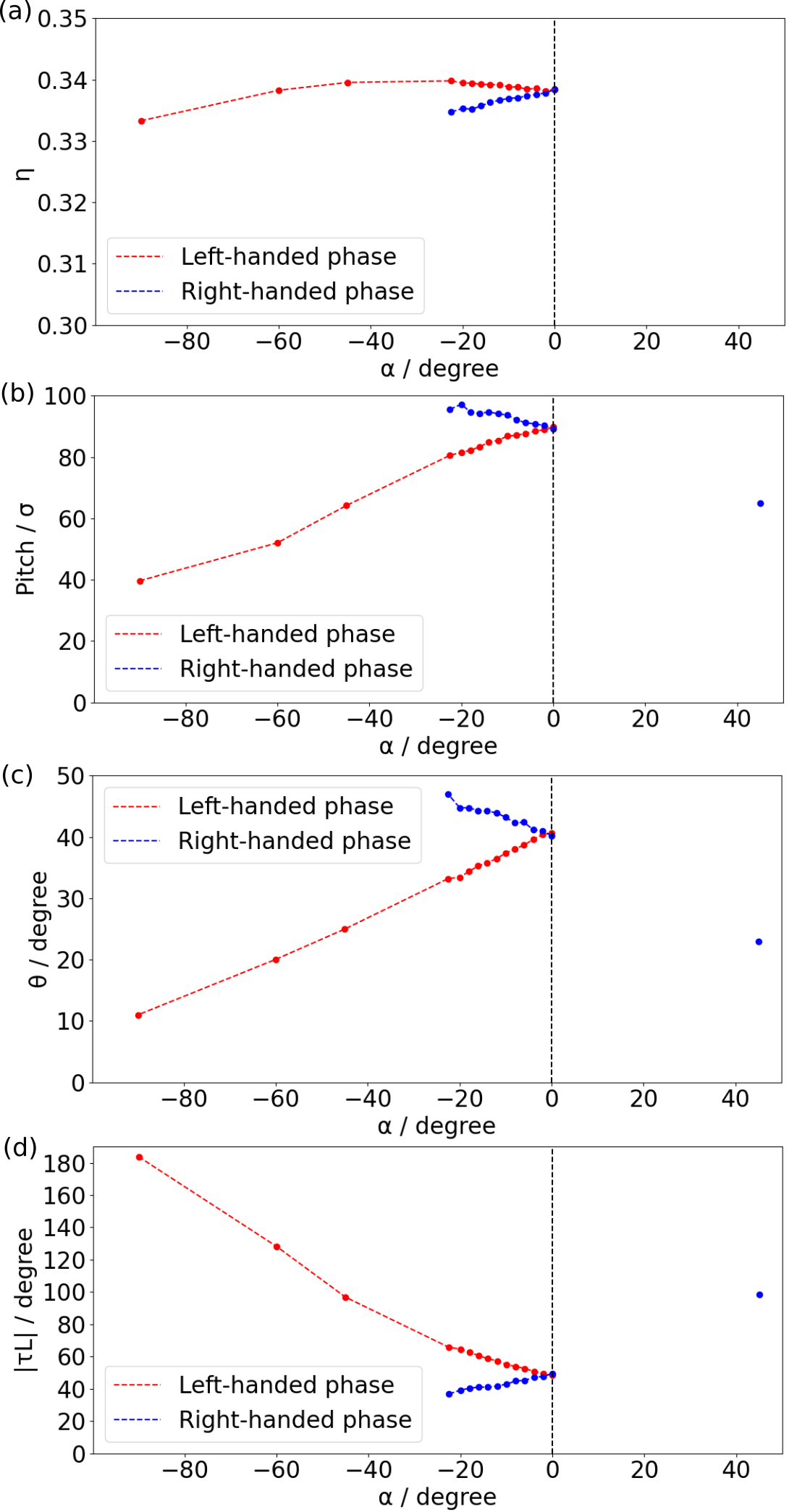}
\caption{\label{fig:LRprops} Properties of the left- and right-handed twist-bend phase as a function of $\alpha$ at pressure $P$=$0.05\,\epsilon\sigma^{-3}$: 
(a) packing fraction $\eta$, (b) pitch $\mathcal{P}$ and (c) cone angle $\theta$ and (d) $\tau L$, where $L=21\sigma$ is the length of the particle.
}
\end{figure}

Firstly, a small difference in the packing fraction between the left- and right-handed phase develops with the left-handed particles able to pack more efficiently, and achieve a higher density for the given pressure, in the left-handed phase (Fig.\ \ref{fig:LRprops}(a)). For example, at the last $\alpha$ value at which a right-handed phase is observed, the left-handed phase has a packing fraction that is 1.5\%  greater. The packing fraction in the left-handed phase is also initially higher than that at $\alpha$=0 indicative of the chirality stabilizing the twist-bend phase. However, at larger $\alpha$ the packing fraction in the left-handed phase begins to decrease. There seems to be an optimal amount of chirality at which the particle chirality best matches the phase chirality and beyond which increasing the chirality begins to hinder the particles ability to pack in the twist-bend phase. A similar picture is seen in the potential energy of the left- and right-handed phases (Fig.\ S2). The left-handed phase has a lower potential energy because the better packing of the particles means less particle overlaps and repulsions despite a higher packing fraction.

Secondly, the geometric properties of the twist-bend nematic begin to depend on the handedness of the phase. In particular, the pitch is shorter and the cone angle smaller for the more stable left-handed phase (Fig.\ \ref{fig:LRprops}(b) and (c)). Both these changes lead to an increase in the torsion of the nematic director field lines (Eq.\ \ref{eq:torsion} and Fig.\ \ref{fig:LRprops}(d)). It is facilitated by the chiral particle being able to better follow the left-handed helical path of the director field. This feature is also evident from the greater deviation between the contours of the achiral particles and the helical path of a director field line representative of the $\alpha$=$-90^\circ$ system than of the $\alpha$=0 system in Fig.\ \ref{fig:TB_schematic}(b). By contrast, the curvature of the field lines is relatively constant as the effect of the decrease in pitch is roughly offset by the effect of the decrease in angle (Eq.\ \ref{eq:curvature}). Contrariwise the left-handed particles are less able to follow the director field in the right-handed phase, thus causing the torsion of the field lines to decrease.

At large $\alpha$, the torsion increases approximately linearly with $\alpha$ (Fig.\ \ref{fig:LRprops}(d)). In this regime, the torsion of the field lines is determined by the particle twist. $\tau L$ represents the net twist of the binormal to the field lines over a particle length and is approximately $2\alpha$ at large $\alpha$. The factor of two is because applying a twist at the centre of the particle has a greater effect on the overall chiral shape of the particle than if the same twist is applied through a constant torsion along the whole particle contour (as is the case for a helix). 

One can also calculate the effect of $\alpha$ on the magnitude of the twist and bend deformation in the twist-bend phase.
The magnitude of the twist deformation is $|\mathbf{n} \cdot \left(\nabla\times\mathbf{n}\right)|= 2 \pi \sin^2\theta/\mathcal{P}=2\pi/\mathcal{P}-\tau$
and the bend is
$|\mathbf{n}\times\left(\nabla\times\mathbf{n}\right)|= 
2\pi \sin\theta\cos\theta/\mathcal{P}=\kappa$ (the square of both these terms enter into Eq.\ \ref{eq:frank}). Note the magnitude of the twist deformation behaves in an opposite manner to the torsion of the field lines. Thus, perhaps counter-intuitively, the introduction of particle chirality leads to a significant reduction in the magnitude of the twist deformation in the preferred chiral twist-bend phase. By contrast, the magnitude of the bend deformation only changes modestly, initially increasing slightly as the magnitude of $\alpha$ increases. 

Overall, this is likely to lead to a reduction in the free-energy cost of forming the twist-bend phase, the uncertainty being that we do not know the values of the elastic constants in Eq.\ \ref{eq:frank} or the contributions of other terms that may be relevant for our low-symmetry particles.\cite{Xu20} The particle chirality reduces the free-energetic costs of forming a twist-bend phase with a higher torsion of the director field lines allowing a twist-bend phase to form with a similar amount of bend but lower twist.

There have been two theoretical approaches applied to the chiral twist-bend phase.\cite{Meyer16,Longa18} In both cases the only term in the free-energy expansion that depends on the particle chirality was a term equivalent to the chiral strength term in Eq.\ \ref{eq:frank}. An immediate consequence of this is that the sign of this term controls the handedness of both the cholesteric and twist-bend phases; i.e.\ they are predicted to have the same handedness.

However, for the particles considered here, as they are weakly twisted, the expectation, confirmed in the next section, is that the handedness of the cholesteric phase will be opposite to the particle chirality\cite{Tortora17,Grelet24} and, thus, opposite to that of the twist-bend phase. Thus, the origins of the chirality transfer in the current twist-bend phases must instead involve chiral terms that were not included in those studies. 

\begin{figure*}[t]
\centering
\includegraphics[width=0.95\textwidth]{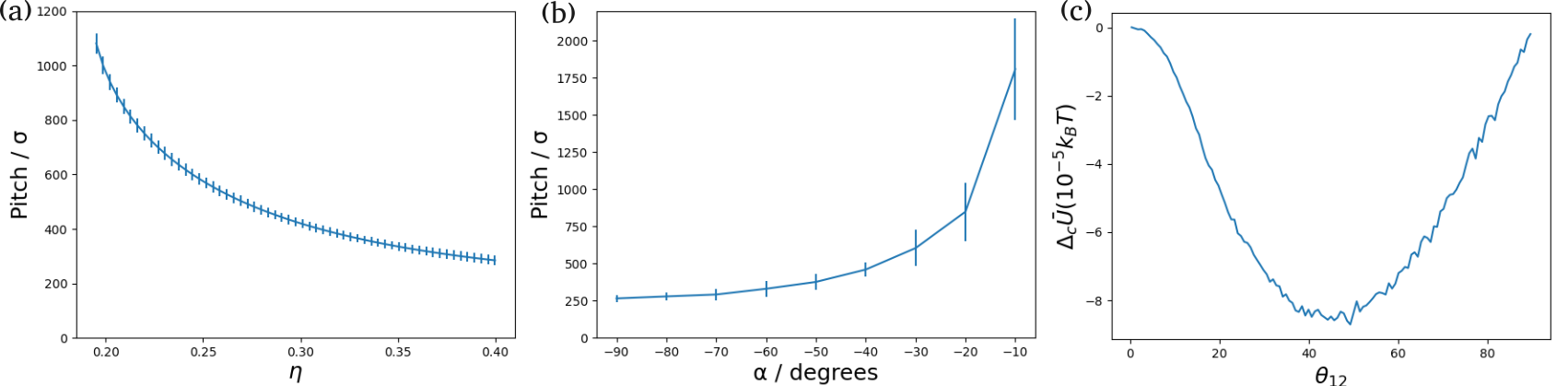}
\caption{\label{fig:DFT} 
The pitch of the cholesteric phase predicted by classical density functional theory 
(a) as a function of packing fraction at
$\alpha$=$-45^\circ$, and
(b) as a function of $\alpha$ at a packing fraction of 0.3. 
(c) Chiral two-body potential of mean force at $\alpha$=$-45^\circ$. 
}
\end{figure*}

Ref.\ \onlinecite{Revignas24} provides an interesting example of a theoretical prediction (using a Maier-Saupe-type approach) of two chiral liquid-crystalline phases with opposite handedness. The free-energy landscape for a right-handed helical particle has two minima, namely a left-handed cholesteric and a screw-like nematic phase, \cite{Kolli14} where the director defined by the internal bonds of the particle (rather than the molecular axis) has right-handed heliconical order that matches the pitch and groove angle of the particle helix. Such a theoretical approach would likely be able to predict opposite phase handedness for particles with the same shape as those considered here, albeit driven by the Maier-Saupe energy term that favours alignment of the particle bonds with the local nematic director, rather than by the packing effects that are relevant to lyotropic systems.

\subsection{The cholesteric phase}
\label{sect:chol}

As noted already, our expectation, based on Straley's geometric argument\cite{Straley76} and previous results for helical particles,\cite{Tortora17,Grelet24} is that the weakly-twisted chiral particles considered here will give rise to a cholesteric phase with opposite-handedness to the particles. This is confirmed by our classical density functional theory calculations. For particles with a left-handed twist, i.e.\ $\alpha< 0$, the cholesteric phase is always predicted to be right-handed.

Fig.\ \ref{fig:DFT} reports the results of our calculations.
Unsurprisingly, as the particles become more twisted the cholesteric pitch becomes shorter (Fig.\ \ref{fig:DFT}(b)).
The magnitude of the pitch also decreases monotonically with packing fraction (Fig.\ \ref{fig:DFT}(a)). Although $k_t$ and $K_2$ both increase with packing fraction, $k_t$ increases more rapidly (Fig.\ S3).
This behaviour of the pitch is in line with previous results for weakly-twisted helical particles where the particle pitch is significantly longer than the particle length.\cite{Tortora17,Grelet24} By contrast, for more strongly-twisted helical particles an increase in the pitch is often seen at higher packing fractions because the increase in nematic order suppresses the pair configurations with a larger interparticle twist that play a key role in chirality transfer at lower packing fractions.\cite{Tortora17} 

To characterize more fully the origins of the preference for a cholesteric phase of opposite handedness to that of the particles, we compute the chiral two-body potential of mean force introduced in Refs.\ \onlinecite{Tortora20,Grelet24}. It is defined as 
\begin{eqnarray}
  \label{eq:chiral_mean}
  \Delta_c \overline{U}(\theta_{12}) &=& \overline{U}_R(\theta_{12}) - \overline{U}_L(\theta_{12}) \\
  &=& k_\mathrm{B} T \log \frac{\big\langle e^{-\beta V_{12}} \big \rangle_L^{(\theta_{12})}}{\big\langle e^{-\beta V_{12}} \big \rangle_R^{(\theta_{12})}}.
\end{eqnarray}
where 
$V_{12}$ is the potential energy of interaction between the pair of particles and
the equilibrium averages are over all pair configurations with a given value of the inter-particle twist angle $\theta_{12}$ and handedness. 
$\theta_{12}=\cos^{-1}\left(\hat{\mathbf{a}}_1\cdot\hat{\mathbf{a}}_2 \right)$ and the handedness of a pair configuration is determined by the sign of $\mathbf{r}_{12} \cdot (\hat{\mathbf{a}}_1 \times \hat{\mathbf{a}}_2)$, where $\mathbf{r}_{12}$ is the centre-of-mass separation.
$\overline{U}_R(\theta_{12})$ and $\overline{U}_L(\theta_{12})$ are angular potentials of mean force associated with right- and left-handed two-particle configurations. Thus, if $\Delta_c \overline{U}(\theta_{12})<0$ right-handed configurations are thermodynamically favoured at that inter-particle twist angle.
$\Delta_c \overline{U}(\theta_{12})$ is depicted in Fig.\ \ref{fig:DFT}(c) for a left-handed particle with $\alpha=-45^\circ$. It is always negative between 0 and 90$^\circ$ 
and has a broad minimum near to $\theta_{12}$=45$^\circ$, thus indicating right-handed configurations are favoured for any twist angle. 
Similar broad minima favouring opposite-handed pair configurations have been seen for other weakly-twisted helical particles where the particle pitch is significantly longer than the particle length.\cite{Grelet24}

\begin{figure}
\includegraphics[width=7.4cm]{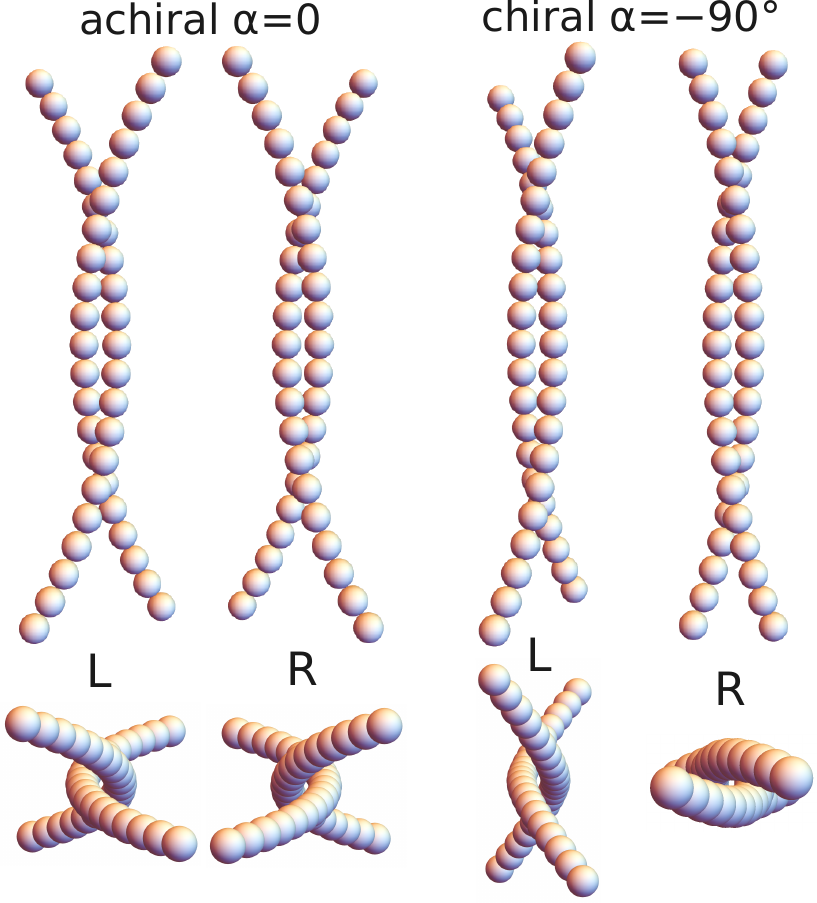}
\caption{\label{fig:pair} Side and top views of left- (L) and right-handed (R) ``anti-polar'' pair configurations with the particle $\mathbf{\hat{b}}$ axes anti-aligned and pointing towards each other. The central spheres of each particle are separated by $2^{1/6}\sigma$ and the left- and right-handed interparticle twists correspond to the minimum required to remove all repulsive interactions. The values of the interparticle twists are $\mp 20.14^\circ$ for the achiral particles, and $-32.6^\circ$ and $6.36^\circ$ for the left-handed chiral particles with $\alpha=-90^\circ$. A negative sign indicates a left-handed interparticle twist and zero corresponds to the long axes (defined in terms of the relevant eigenvector of the gyration tensor) of the two particles being aligned.
}
\end{figure}

``Anti-polar'' pair configurations in which the $\mathbf{\hat{b}}$ particle axes point towards each other (Fig.\ \ref{fig:pair}) are likely to make a significant contribution to chirality transfer in the cholesteric. If the long axes are also aligned then the particles will overlap as they approach each other. Hence, there is a natural tendency for the particles to twist. For achiral particles left- and right-twisted configurations are equally likely. However, for the chiral particles the twist required to remove overlaps is smaller when the pair configurations have an opposite handedness to the particle. The particles naturally wrap around each other in these configurations (right-hand image in Fig.\ \ref{fig:pair}).

As the packing fraction increases and the twist-bend phase boundary is approached, there is likely to be an increase in polar correlations and a concomitant decrease in the contribution from the anti-polar configurations highlighted above. Thus, the uniaxial assumption in the current density functional theory approach may well contribute to an underestimation of the pitch at higher packing fractions.
Furthermore, the polar order present in the twist-bend nematic makes these anti-polar pair configurations even less relevant for the twist-bend phase and helps to explain why the sign of $k_t$ is not the dominant determinant of the preferred chirality of the twist-bend phase. 
By contrast, if particle chirality were introduced in a way that polar pair configurations were the most significant contributors to chirality transfer in the cholesteric phase, it might then be that the cholesteric and twist-bend phase would have the same chirality. 

\subsection{Chiral dopants}
\label{sect:doped}

\begin{figure}
\includegraphics[width=8.4cm]{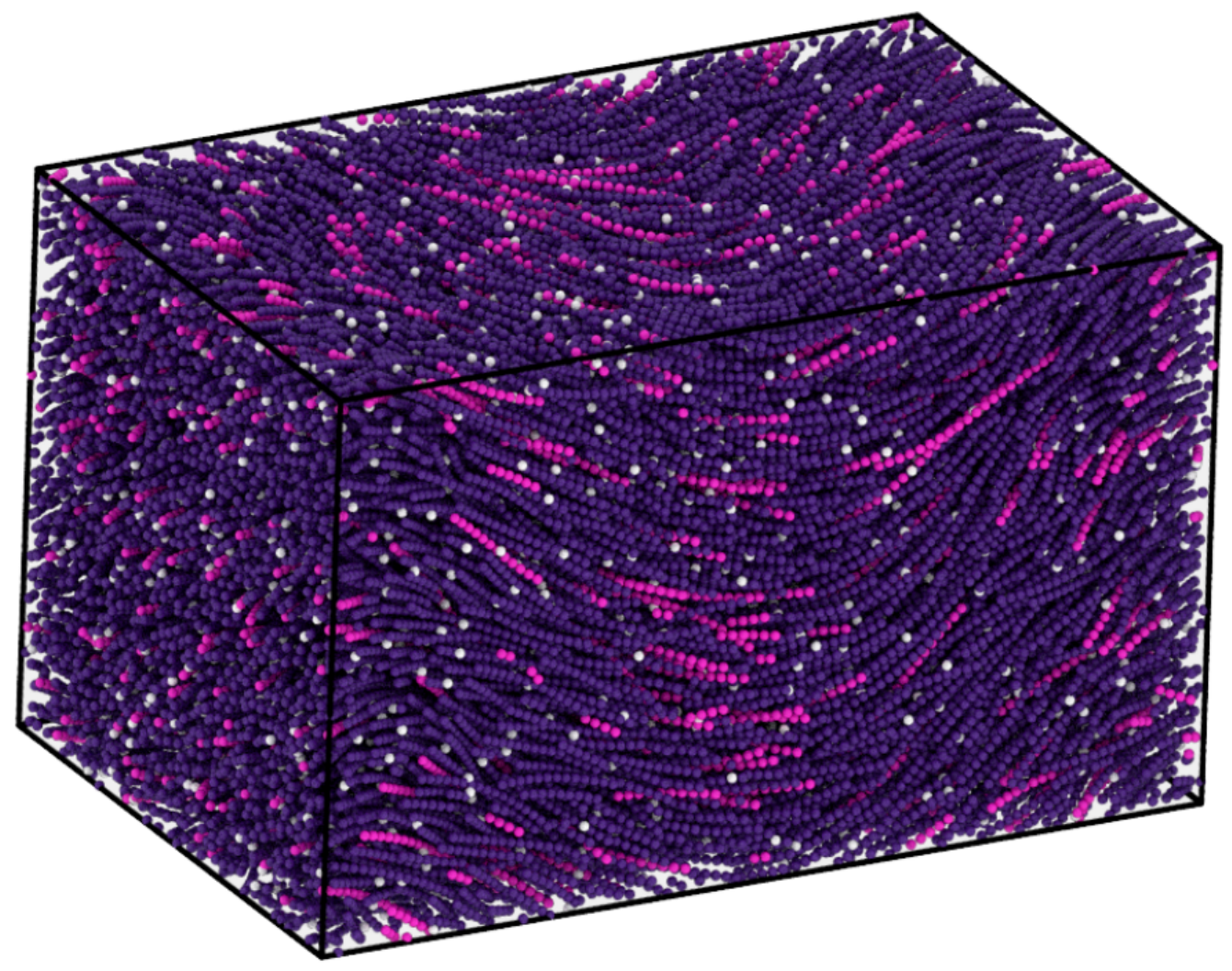}
\caption{\label{fig:doped} Snapshot of a left-handed twist-bend phase in a system of achiral particles (dark purple) doped with 10\% of chiral $\alpha$=$-90^\circ$ particles (pink).
}
\end{figure}

We also considered the effects of adding chiral dopants to an achiral system. Specifically, we added a small fraction of left-handed mesogens to a system of otherwise achiral mesogens (Fig.\ \ref{fig:doped}). The choice of which mesogens were chiral in the initial lattice configuration was random. The effects on the handedness of the twist-bend nematic phase are shown in Table \ref{tab:doping}.

As expected from the above results for one-component systems, left-handed dopants led to the formation of a left-handed phase being favoured. This effect increases both with increasing $|\alpha|$ and dopant concentration with the addition of 10\% of chiral mesogens with $\alpha$=$-90^\circ$ sufficient to cause the left-handed phase to be always observed. The dopants add a bias to the free-energy landscape making it more favourable for a phase to form with chiral environments that match the handedness of the dopant. 

\begin{table}
  \centering
    \begin{ruledtabular}
    \begin{tabular}{ccc}
    [Dopant] & \multicolumn{2}{c}{LH:RH} \\
    \cline{2-3}
    & $\alpha$=$-$45\textdegree & $\alpha$=$-$90\textdegree \\
    \hline 
    5\% & 61\%:39\% & 90\%:10\%  \\
    10\% & 88\%:12\% & 100\%:0\%  \\
    \end{tabular}
    \end{ruledtabular}
  \caption{\label{tab:doping}Fraction of simulations 
  that end up in the left- or right-handed twist-bend nematic phase for doped systems. Different dopant concentrations and $\alpha$ for the dopant particles were considered. Pressure $P$=$0.04\,\epsilon\sigma^{-3}$.
  Statistics were based on a total of 20 runs.  
  }
\end{table}

Differences in packing fraction between the left- and right-handed phases are less apparent than for the one-component systems. Only for  5\% doping with $\alpha$=$-90^\circ$ particles does the left-handed phase have a clearly higher packing fraction (and a lower potential energy) than the right-handed phase. In this case, the packing fractions are both lower than for the achiral system. 
By contrast, doping with the $\alpha$=$-45^\circ$ particles does not lead to any significant changes in the packing fraction compared to the achiral system with the differences in the packing fraction of the right- and left-handed phases being within statistical error.
Thus, there is no clear stabilization of the twist-bend phase by doping and only for doping with the $\alpha$=$-90^\circ$ particles 
were we able to observe a clear thermodynamic component to the bias favouring the left-handed phase. 
These features are perhaps unsurprising as there is likely to be less compatibility between the shapes of the chiral and achiral particles when they pack together in polar configurations.

Similarly, the geometric properties of the left- and right-handed twist-bend phases are less different than in the case of the one-component systems. These properties are likely dominated by what is optimal for the majority achiral particles rather than the dopants. For example, for 5\% doping with $\alpha$=$-45^\circ$ mesogens the left-handed phase only has a 3.5\% smaller pitch than the right-handed phase.

\section{Conclusions}
\label{sect:conc}

Here, we have investigated the phase behaviour of lyotropic chiral bent rods using a mixture of simulations and theory, particularly focussing on the chirality of the nematic phases that result. For our weakly twisted particles, density functional theory predicts that the cholesteric phase has a handedness that is opposite to that of the particles, in agreement with Straley's simple geometric argument.\cite{Straley76} By contrast, the preferred handedness of the twist-bend phase is the same as that of the particles. 
That the same chiral particle can lead to phases of different handedness may seem somewhat counterintuitive. 
However, it highlights the importance of the nature of a phase's order (e.g.\ the director field) for understanding the effects of particle chirality. 

Our proposed explanation for the twist-bend phase having the same handedness as the particle is that it then allows the local particle contour to better follow the helical director field lines. This in turn allows the particles to pack more efficiently in the twist-bend phase. By contrast, when the particles have opposite handedness, they are less able to follow the director field and pack less efficiently. This explanation also provides an understanding of how the properties of the right- and left-handed twist-bend phases differ in the presence of particle chirality. The match between the phase and particle chirality facilitates a greater torsion of the director field lines; as the curvature of the director field lines remains nearly constant (as they approximately match the particle curvature) this is achieved by a decrease in both the cone angle and the pitch of the twist-bend phase. Perhaps surprisingly, the net effect of the particle chirality is, hence, to reduce the magnitude of the twist deformation in the twist-bend phase.

One of the interesting features of our results is that the packing fraction initially increases on the introduction of chirality reaching a maximal value at $|\alpha|\approx 30$. Thus, chirality stabilizes the twist-bend phase as is also evident in the phase diagram.
For achiral particles, reaching an optimal torsion is resisted by the inability of the achiral particles to sufficiently follow the director field lines as the torsion becomes larger. By contrast, for highly twisted particles, the particle twist enforces a higher torsion to the director field lines than is conducive to optimal packing.

We should emphasize that the trends that we have observed here reflect the specific way that we have introduced chirality into the bent particles, namely through the introduction of a twist at the centre of the particles. Although we expect broadly similar results for alternative ways of introducing twist, e.g.\ continuously along the length of the particle as for particles derived from a helix, there are many other ways of breaking the mirror symmetry of the $C_{2v}$ achiral bent particles and one should not necessarily expect a uniform response to the introduction of chirality. That is one of the reasons why we have not sought to compare to the available results on thermotropic chiral twist-bend phases for which the mesogens and their interactions are generally more complex and varied than for the lyotropic particle that we consider here.

The current results also present a challenge to theory as the previous theoretical descriptions of the chiral twist-bend nematic have both predicted that the cholesteric and twist-bend phases would have the same handedness.\cite{Meyer16,Longa18} This prediction stems from the introduction of just a single chiral term in the free energy that determines the handedness of both phases. Clearly, additional chiral terms are needed to describe the chirality transfer in a twist-bend phase of the current particles.
Although the complete set of order parameters and terms in a Landau-de Gennes free-energy expansion for particles with $C_2$ symmetry have been outlined,\cite{Xu20} the low symmetry means that the number is large. Simulations could perhaps play a role in narrowing that number down to those that are in practice relevant for the chiral twist-bend nematic.

It would also be interesting to study further the behaviour of the cholesteric phase
as the phase boundary with the twist-bend phase is approached; in particular, how increasing polar correlations and the decrease of the bend elastic constant affect the twist of the cholesteric phase. This would require significantly larger simulations than used here and perhaps boundary conditions that aid cholesteric formation.\cite{Dussi16} Interesting pre-transition effects have been seen in experiments.\cite{Walker19,Ozegovic24b} 

Although the main focus of this paper has been on understanding some of the fundamental mechanisms of chirality transfer in twist-bend phases, it may well be possible to realize equivalents of the particles considered here using DNA origami. The cholesteric phases of twisted straight DNA origami nanotubes have been studied previously.\cite{Siavashpouri17,Tortora20} A series of bent DNA origami rods with differing degrees of twist should be simple to design, as it is well understood how to introduce bend and twist into DNA origami.\cite{Dietz09} 
The DNA origami would also be inherently chiral due to the chirality of DNA at the molecular level. Although DNA origami are relatively stiff, thermal shape fluctuations may also be relevant.\cite{Tortora20}

\section*{Supplementary Material}

The supplementary material details further results from simulations and the density functional theory, as well as a brief review of the properties of helical curves.

\begin{acknowledgments}
H.C. would like to thank Dr. Massimiliano Chiappini (Utrecht University) for helpful discussions. H.C. acknowledges support from the Beckman Fellowship, UIUC. A.E.M.\ is grateful for financial support from the Clarendon Fund, University of Oxford.
\end{acknowledgments}

\section*{Author Declarations}

\subsection*{Conflict of Interest}
The authors have no conflicts to disclose.

\section*{Data Availability}
The code used to perform the classical density functional calculations is available at \url{https://github.com/mtortora/chiralDFT}. Data associated with the project will be made available at the Oxford University Research Archive (ORA).

%

\end{document}


\title{Supplementary Material: Chirality transfer in lyotropic twist-bend nematics}

\author{Anna Ashkinazi}
\affiliation{Physical \& Theoretical Chemistry Laboratory, Department of Chemisty, University of Oxford, South Parks Road, Oxford OX1 3QZ, United Kingdom}
\author{Hemani Chhabra}%
\affiliation{Beckman Institute, University of Illinois Urbana-Champaign, IL, USA}
\author{Anouar El Moumane}
\affiliation{Physical \& Theoretical Chemistry Laboratory, Department of Chemisty, University of Oxford, South Parks Road, Oxford OX1 3QZ, United Kingdom}
\author{Maxime M.\ C.\ Tortora}
\affiliation{Department of Quantitative and Computational Biology, University of Southern California, Los Angeles, CA, USA}
\author{Jonathan P.\ K.\ Doye}
\email{jonathan.doye@chem.ox.ac.uk}
\affiliation{Physical \& Theoretical Chemistry Laboratory, Department of Chemisty, University of Oxford, South Parks Road, Oxford OX1 3QZ, United Kingdom}

\date{\today}


\maketitle 

\setcounter{figure}{0}
 \makeatletter
 \renewcommand{\thefigure}{S\@arabic\c@figure}
 \setcounter{equation}{0}
 \renewcommand{\theequation}{S\@arabic\c@equation}
 \setcounter{table}{0}
 \renewcommand{\thetable}{S\@arabic\c@table}
 \setcounter{section}{0}
 \renewcommand{\thesection}{S\@arabic\c@section}

\section{Further results}

\begin{figure}
\includegraphics[width=8.4cm]{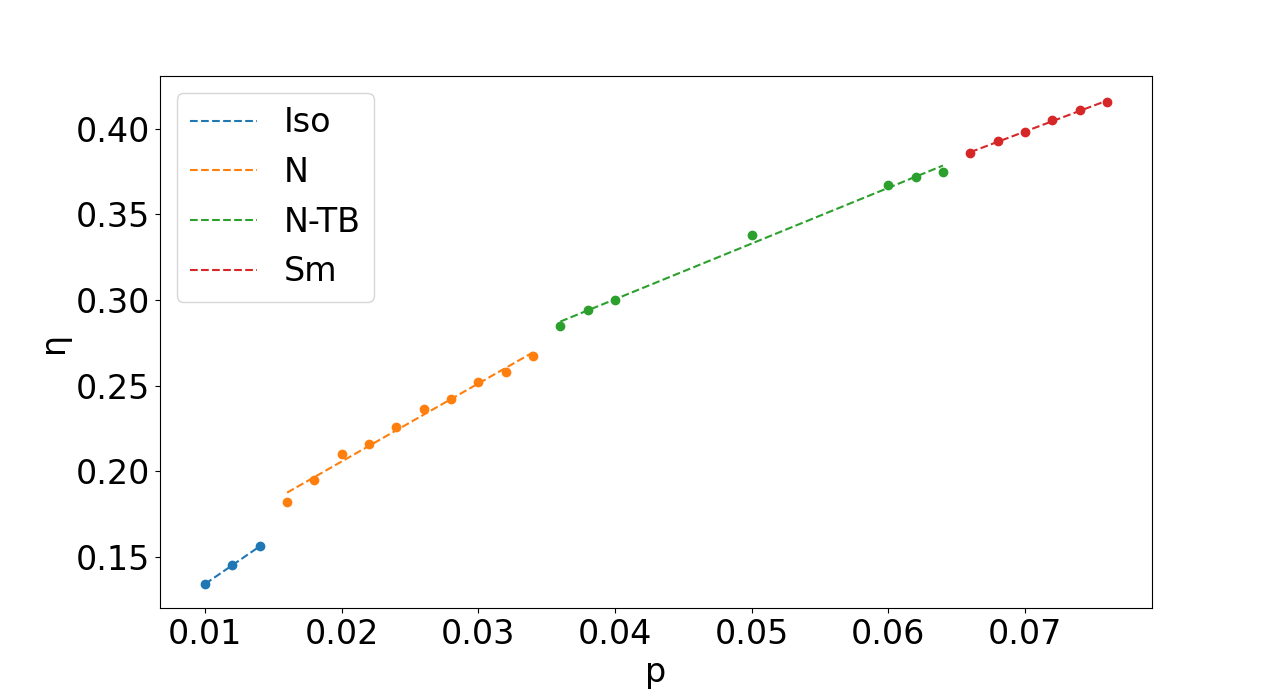}
\caption{\label{fig:eofs} Equation of state for mesogens with $\alpha$=0. The points are coloured by the observed phase and the lines are linear fits to the data in each region.
The transitions from the isotropic to nematic phase and nematic to twist-bend nematic involve clear changes of slope and a jump in the packing fraction. 
}
\end{figure}

Fig.\ \ref{fig:eofs} shows the equation of state for the achiral bent particles. The changes of slope and differences in packing fraction between phases are sufficient to identify the transitions between the isotropic, nematic, nematic twist-bend and smectic phases as the pressure is increased. The changes in the relevant order parameters provide further confirmation of the phase changes.

\begin{figure}
\includegraphics[width=8.4cm]{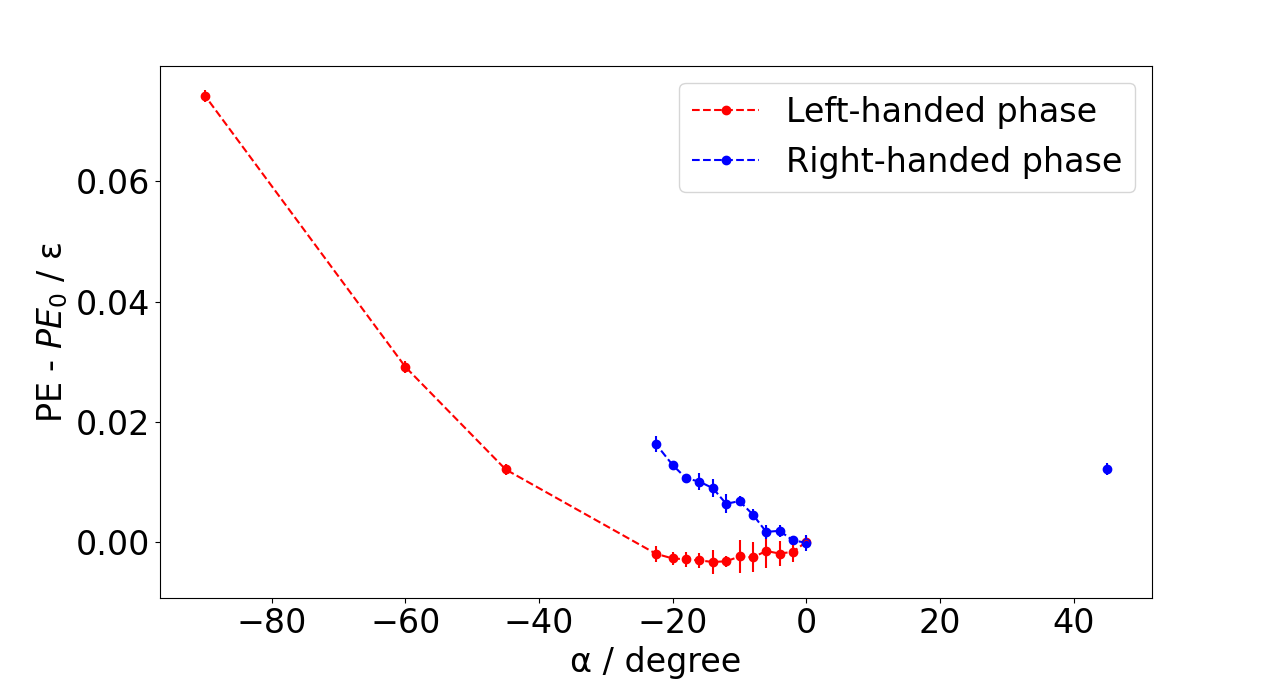}
\caption{\label{fig:Ehand} Potential energy of the twist-bend phase at $P$=$0.05\,\epsilon\sigma^{-3}$ (relative to $\alpha$=0) as a function of $\alpha$ for the left- and right-handed phases. 
}
\end{figure}

Fig.\ \ref{fig:Ehand} shows the potential energy of the left- and right-handed twist-bend phases as a function of $\alpha$ when the pressure is kept constant. Although the potential energy is not a direct measure of stability (at constant pressure that is the Gibbs free energy), as a contributor to that free energy, it can provide useful insights. Moving away from $\alpha=0$ to negative values, the potential energy of the left-handed phase decreases. The inter-particle repulsions are reduced due to the better packing of the left-handed particles into a left-handed phase, this despite an increase in packing fraction for the left-handed phase. By contrast, the potential energy of the right-handed phase is uniformly increasing. The potential energy of the left-handed phase reaches a minimum about $\alpha=-15^\circ$ and then increases as $|\alpha|$ increases further. This behaviour is similar to that for the packing fraction (Fig.\ 7(a) in the main text), albeit with a turning point at smaller $|\alpha|$ due to the effects of the changes in the packing fraction.

\begin{figure}
\includegraphics[width=8.4cm]{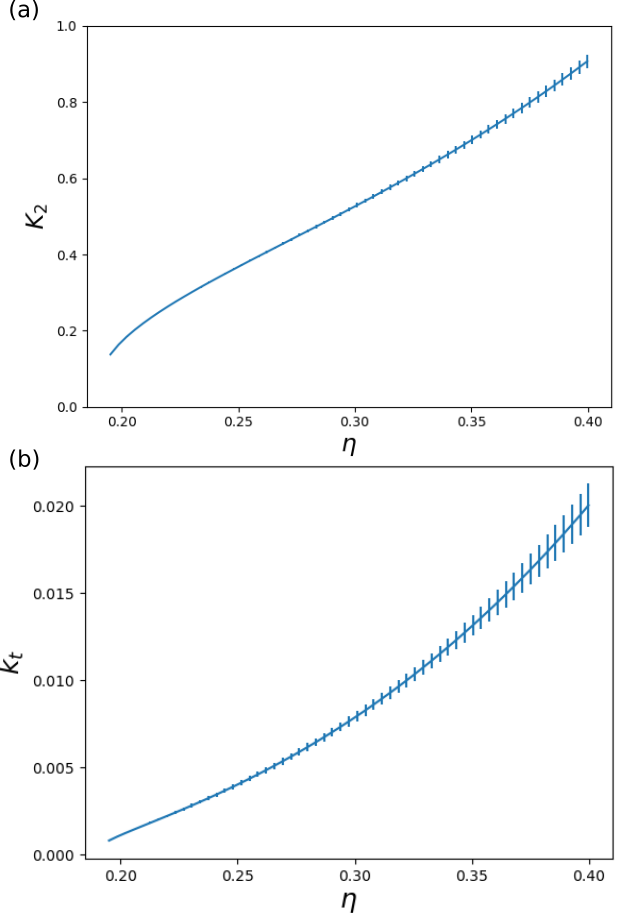}
\caption{\label{fig:dft_elastic} The twist elastic constant $K_2$ and the chiral strength $k_t$ predicted by classical density functional theory as a function of packing fraction $\eta$. Although both increase with packing fraction, $k_t$ increases more rapidly leading to a monotonically decreasing pitch (Fig.\ 8).
}
\end{figure}

Fig.\ \ref{fig:dft_elastic} shows how the twist elastic constant $K_2$ and the chiral strength $k_t$, as calculated using classical density functional theory, vary with packing fraction for a chiral particle with $\alpha=-45^\circ$. Firstly, $k_t$ is positive indicating a tendency for these left-handed particles to form a right-handed phase. Both quantities increase with packing fraction. The uniform increase in $K_2$ is simply due to the increased density and reduced distance between particles increasing the cost of deformations due to a greater likelihood of the deformations leading to increased inter-particle repulsions. The increase in $k_t$ is because as the particles get closer they feel each other's chirality more and there is a greater driving force for inter-particle twist in order to optimize local packing.

As the cholesteric pitch is given by $2\pi K_2/k_t$ the effect on the pitch is determined by the relative magnitude of the increases. In this case $k_t$ increases more rapidly and so the cholesteric pitch is predicted to uniformly decrease (Fig.\ 8(a)) in the main text).

\section{Properties of helices}

Here we review some of the properties of helices, in particular, those that lead to Eqs.\ 7 and 8 in the main text that describe the torsion and curvature of a helix.

A helix can be defined parametrically as
\begin{equation}
\mathbf{r}=
\begin{pmatrix}
a \cos t \\
a \sin t \\
bt
\end{pmatrix}
\end{equation}
where $a$ is the radius of the helix and the pitch $\mathcal{P}=2\pi b$. 

The curvature $\kappa$ and torsion $\tau$ are defined through the Frenet-Serret equations.
\begin{eqnarray}
\frac{d\mathbf{T}}{ds}&=&\kappa\mathbf{N} \\
\frac{d\mathbf{N}}{ds}&=&-\kappa\mathbf{T}+\tau\mathbf{B} \\
\frac{d\mathbf{B}}{ds}&=&-\tau\mathbf{N}
\end{eqnarray}
where $\mathbf{T}$ is the tangent, $\mathbf{N}$ is the normal and $\mathbf{B}$ is the binormal. For a helix
\begin{equation}
\mathbf{T}=\frac{1}{\sqrt{a^2+b^2}}
\begin{pmatrix}
-a \sin t \\
a \cos t \\
b
\end{pmatrix},
\end{equation}
\begin{equation}
\mathbf{N}=
\begin{pmatrix}
-\cos t \\
\sin t \\
0
\end{pmatrix}
\end{equation}
and
\begin{equation}
\mathbf{B}=\frac{1}{\sqrt{a^2+b^2}}
\begin{pmatrix}
b \sin t \\
-b \cos t \\
a
\end{pmatrix}
\end{equation}
Thus, 
\begin{equation}
\kappa=\frac{a}{a^2+b^2}
\end{equation}
and
\begin{equation}
\tau=\frac{b}{a^2+b^2}
\end{equation}
These can be rewritten in terms of the pitch and cone angle giving the expressions in the main text 
(Eqs.\ 7 and 8)
using $\tan\theta=a/b$ where geometrically $\theta$ is the angle between the tangent and the helix axis.

Although for our model chiral particle, we used a bent particle with a single twist at the centre of the particle, we could have also used alternative chiral particles that were sections of a helix. In particular, one could have studied helical particle with constant curvature $\kappa$, but where one varied the torsion $\tau$. $\tau=0$ would then correspond to the achiral limit.

The above equations can be rearranged to give the helical parameters $a$ and $b$ in terms of $\kappa$ and $\tau$:
\begin{equation}
a=\frac{1}{\kappa\left(1+\left(\tau/\kappa\right)^2\right)}
\end{equation}
\begin{equation}
b=\frac{1}{\tau\left(1+\left(\kappa/\tau\right)^2\right)}
\end{equation}
$\kappa$ is set by that for the achiral particle. i.e.\
\begin{equation}
\kappa L=\theta_\mathrm{central}=\frac{\pi}{4}
\end{equation}
